\documentclass[]{aastex631}
\usepackage{graphicx}
\usepackage{xspace}
\usepackage{txfonts}
\usepackage[utf8]{inputenc}

\newcommand{\ipole}{\texttt{ipole}\xspace}

\begin{document}

\title{Prospects for ray-tracing light intensity and polarization in  models of accreting compact objects using a GPU}

\correspondingauthor{Monika A. Moscibrodzka}
\email{m.moscibrodzka@astro.ru.nl}

\author[0000-0002-4661-6332]{Monika A. Moscibrodzka}
\affiliation{Department of Astrophysics/IMAPP, Radboud University,P.O. Box
 9010, 6500 GL Nijmegen, The Netherlands}

\author[0000-0002-3244-7072]{Aristomenis I. Yfantis}
\affiliation{Department of Astrophysics/IMAPP, Radboud University,P.O. Box
 9010, 6500 GL Nijmegen, The Netherlands}



\begin{abstract}
The Event Horizon Telescope (EHT) has recently released high-resolution images of accretion flows onto two supermassive black holes. Our physical understanding of these images depends on accuracy and precision of numerical models of plasma and radiation around compact objects. The goal of this work is to speed up radiative-transfer simulations used to create mock images of black holes for comparison with the EHT observations. A ray-tracing code for general relativistic and fully polarized radiative transfer through plasma in strong gravity is ported onto a graphics processing unit (GPU). We describe our GPU implementation and carry out speedup tests using models of optically thin advection-dominated accretion flow (ADAF) onto a black hole realised semi-analytically and in 3D general relativistic magnetohydrodynamics simulations, low and very high image pixel resolutions, and two different sets of CPU+GPUs. We show that a GPU with high double precision computing capability can significantly reduce the image production computational time, with a speedup factor of up to approximately $1200$. The significant speedup facilitates, e.g., dynamic model fitting to the EHT data, including polarimetric data. The method extension may enable studies of emission from plasma with nonthermal particle distribution functions for which accurate approximate synchrotron emissivities are not available. The significant speedup reduces the carbon footprint of the generation of the EHT image libraries by at least an order of magnitude.
\end{abstract}



\section{Introduction}

The Event Horizon Telescope has recently released images of accretion flows in immediate vicinity of black holes (e.g., \citealt{ehtI:2019,ehtI:2022}). Physical understanding of these images requires models of plasma in strong gravity and radiative-transfer simulations that translate the plasma densities, temperatures, velocities and plasma magnetizations into synchrotron emission intensity maps. To create these synthetic maps of black holes we usually postprocess general relativistic magnetohydrodynamic (GRMHD) simulations of gas falling onto the black hole's event horizon using radiative-transfer schemes. The radiative-transfer results, the images, depend on several parameters and to cover the large parameter space one snapshot of black hole simulation 
can generate a large number of possible model appearances. For example, to interpret images of the Galactic center's black hole, Sgr~A*, several GRMHD simulations were carried out to estimate parameters of the source. Those generated a few million template images \citep{ehtV:2022}. 
The creation of image libraries is therefore computationally expensive. 
Parameter estimation of black holes and their accretion flows can be also carried out by comparing model images to the data using Bayesian parameter estimation networks. In this approach the mock images are generated dynamically for different parameters in long Markov Chain Monte Carlo chains until posterior distribution functions for parameters converge. Finally, black hole images, depending on the observer's frequency, display fine features such as photon rings (e.g., \citealt{gralla:2019}, \citealt{johnson:2020}, \citealt{davelaar:2022}), so making them at higher resolutions is desirable. It is therefore sensible that mock images of black holes are created 
at the highest resolution possible and in as little time as possible. 

\begin{figure*}
\centering
\includegraphics[width=0.45\linewidth]{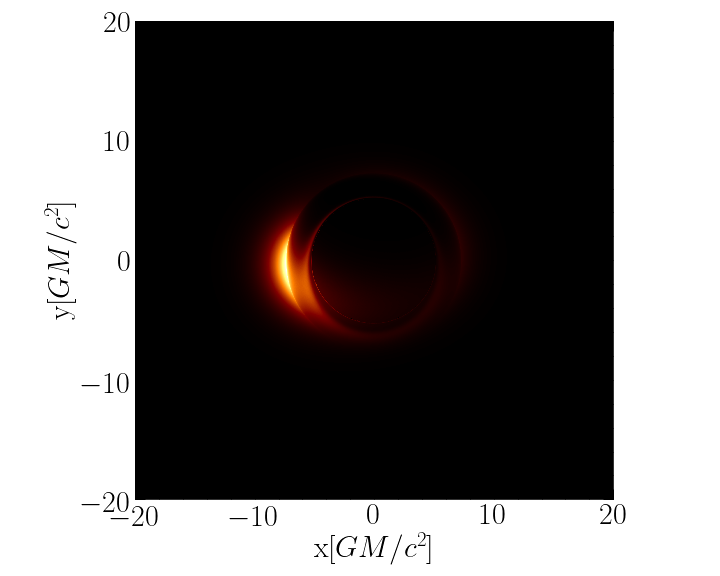}
\includegraphics[width=0.45\linewidth]{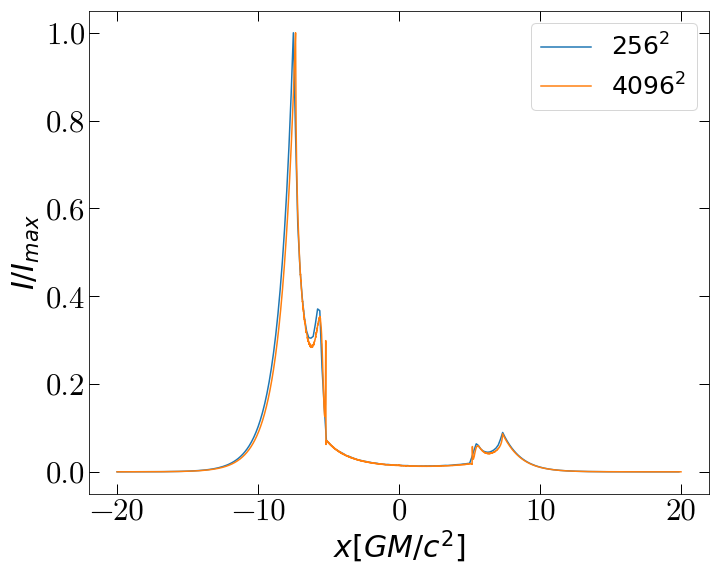}\\
\includegraphics[width=0.45\linewidth]{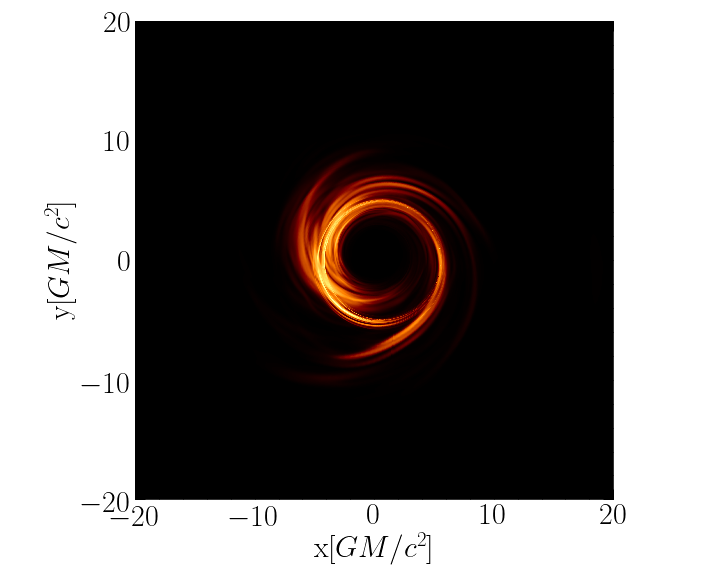}
\includegraphics[width=0.45\linewidth]{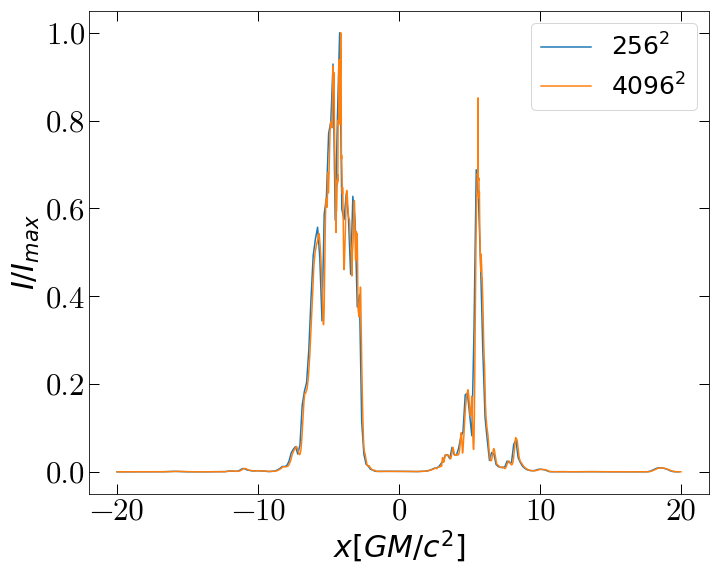}
\caption{Intensity maps and intensity profiles of plasma around a black hole event horizon (the central depression in intensity is the shadow of a black hole). Top panels show a semianalytic model of ADAF seen at a viewing angle of $60^\circ$. 
Lower panels show a model of an accretion flow produced in 3D GRMHD simulations with viewing angle of $20^\circ$. Both images are made at an observer's frequency of 230~GHz which corresponds to the current observing frequency of EHT. The images in the left panels have resolutions of $4096\times 4096$ pixels. The right panels show intensity profiles based on images with high and low resolution. High-resolution images are obviously better at capturing details such as the photon rings.}\label{fig:images_res}
\end{figure*}

Two types of improvements of the ray-tracing schemes have been proposed and developed to make high-resolution imaging simulations feasible. Some codes have been adopted to be executed 
using graphics processing units (GPUs) which are usually faster than central processing units (CPUs). A notable example of such a code is the GRay ray-tracing code developed by \citet{chan:2013} (see also similar efforts by \citealt{pu:2016} and \citealt{davelaar:2018}). The second proposed approach is to carry out adaptive ray-tracing calculations in which the number of rays (or image pixels) is adjusted to the prominent image features. In this approach one can achieve accurate, high-resolution images of accreting object by simulating $\sim10$~times fewer rays than typically (see, e.g., \citealt{gelles:2021,wong:2021}).

In this work we adopt the former approach and rewrite a general relativistic ray-tracing code for GPU execution. Our code \ipole is a ray-tracing code for covariant fully polarized radiative-transfer simulations with main application to model images of black holes and their accretion flows. We show that regardless of the simulated problem complexity, simulations of high-resolution images can be carried out in much shorter time compared to the usual CPU computations or even to the adaptive ray-tracing approach. To our knowledge, this has not been yet demonstrated for fully polarimetric ray-tracing calculations which are far more complex and computationally demanding compared to unpolarized radiative transfer.

The presentation of our numerical experiment is structured as follows. In Section~\ref{sec:method} we describe the CPU code's main features and its GPU extension. In Section~\ref{sec:results} we benchmark the code performance on single CPU and GPU devices. The efficiency of the code and its further applications are discussed in Section~\ref{sec:discussion}.
 

\section{Method}\label{sec:method}

\subsection{CPU code}
We first briefly describe the code to outline the complexity of the radiative-transfer simulations in strong gravity.
\ipole's numerical scheme has been introduced and described in full detail in \citet{moscibrodzka:2018}.
The code has been extensively tested in the original paper and recently also compared to many other polarized ray-tracing schemes (Prather et al. in preparation). The code creates images of arbitrary relativistic plasma and magnetic field configuration provided either by a separate GRMHD simulation or a semianalytic prescription. Radiative processes that \ipole takes into account are synchrotron emission, synchrotron absorption and Faraday effects such as Faraday rotation and Faraday conversion. When modeling images of black holes the code parameters include the mass accretion rate scaling factor, the electron temperature prescription (thermal or nonthermal electron distribution functions), the black hole spin, and the viewing angle of an external observer at a large distance from the black hole. Semianalytic models of plasma have more free parameters, such as the geometry of magnetic fields, prescription for density and magnetic field strength as a function of position and plasma four-velocities which all can be a function of position in space and time.

Mock images of GRMHD models and semianalytic models of accretion can be produced for an arbitrary observing frequency, for a chosen field of view and chosen image resolution. The resolution is usually set to be a power of 2 (typically $256^2$, $512^2$,...) to enable fast Fourier transformations and translation of the images into a visibility function observed by the EHT without any image regridding. Initially the code launches photons from the point of the camera and integrates a null geodesic equation for every single pixel in the camera. The step size of the integration is adaptive and parameterized and the parameter that controls the overall integration step size must be adjusted depending on the physical problem or numerical grid. Once the geodesic integration is over and saved the covariant fully polarized radiative-transfer equations are integrated along the geodesic toward the observer. During this integration the code interpolates the plasma quantities to the point at every single step on a geodesic to evaluate emission, self-absorption and Faraday coefficients at a given point. The speed of this integration is slower for the semianalytic models compared to the GRMHD simulations because in the semianalytic models the plasma properties are calculated along with the geodesic integration  while in the GRMHD  models we read out the plasma quantities from precomputed three 3D arrays. Typically these arrays have approximately $200\times200\times200$ (or more) elements and they are allocated dynamically.

Figure~\ref{fig:images_res} displays images of the two aforementioned models created by the \ipole code at the EHT observing frequency. The semianalytic model used in this work is described in full detailed by \citet{vos:2022} and the GRMHD model is based on models presented by \citet{moscibrodzka:2014}.
The \ipole scheme also follows light polarization (which is not explicitly shown in Figure~\ref{fig:images_res}). The polarization in \ipole is represented in the coordinate frame as a coherency matrix which is a complex quantity (see the method paper for more details: \citealt{moscibrodzka:2018}). Hence during integration the code carries out complex multiplications and additions. 

\ipole code is parallelized using OpenMP directives so it can be run as a multi-thread CPU job. Since the radiative transfer integration along separate light-rays is independent of each other the scaling of the code parallelization is very efficient.


\subsection{GPU implementation}

\ipole was originally written in C language. In this work we extend the code with the Compute Unified Device Architecture (CUDA) language to enable ray-tracing radiative transfer to be run in parallel on Nvidia GPU. The code scheme and philosophy are mostly unchanged except for one detail. The CPU code integrates the geodesics backwards in time (away from the observer) and saves the path to integrate radiative-transfer equation forward in time. The GPU code does this integration in both directions, away from the observer (backwards in time) and toward the observer (forward in time), without saving the path  in both cases.  This is done to limit the memory usage per single GPU thread.  Notice also that our geodesics integrator is not perfectly time-reversible so geodesic integrated forward in time may result in a slightly different path. However the integration is parameterized with an $\epsilon$ and for a typically adopted value of  $\epsilon$ the differences in all Stokes parameter images produced by CPU and GPU codes are negligible. 

The main idea behind CUDA--\ipole is to divide the image pixel grid into blocks assuming that each block is made of $16\times16$ threads. This particular size of CUDA blocks seems the most optimal and feasible for our application. For smaller block sizes the code execution slows down. The larger sizes (with maximum allowed CUDA block of $32\times32$ threads) are restricted to us due to memory usage. To carry out the complex radiative-transfer simulations most of the global and constant variables in the code are copied either into a constant memory of the GPU or managed memory that is visible by CPU and all the GPU blocks/threads. A final small adjustment is done in the part of the code that deals with complex variables, these are handled differently in C and CUDA. We have checked that both original \ipole and CUDA--\ipole produce the same images. 

\section{Results}\label{sec:results}


\begin{table*} \footnotesize
\centering
\begin{tabular}{ c c c  c c c  c c c}  
\hline 
Image & \multicolumn{4}{c}{Semianalytic Model} & \multicolumn{4}{c}{GRMHD Model}\\ 
Resolution  & $CPU_1$ & $GPU_1$ & $CPU_2$ & $GPU_2$ &  $CPU_1$ & $GPU_1$ & $CPU_2$ &$GPU_2$\\
\hline 
$256\times256$  & 650.4 & 1.64 & 338.1 & 6.9 & 332.5 & 0.76 &179.8 & 3.5\\
$512\times512$  & 2589.8 & 3.0 & 1534.9 & 22.6& 1448.4& 1.86 & 806.9& 12.4\\
$1024\times1024$  & 10485.8 & 9.69 & 6418.7 & 83.6& 5343.1 & 6.28 & 3669.7 &46.8 \\
$2048\times2048$  & 43,485.6 & 35.3  & 25,691.8& 325.5 &22,772.6  & 23.3 &14,883.3 & 183.9 \\
$4096\times4096$  & 168,192.1 & 131.5 & 102,956.4 & 1228.8 & 87,200.9& 88.9  & 60,174.3 &728.3 \\
\hline
\end{tabular}
\caption{\ipole's execution times for making one fully polarized (Stokes ${\mathcal I,Q,U,V}$) image of a compact object based on a Semianalytic Model and One Time Slice of a GRMHD simulation. All times are the wall-clock times in units of seconds. Results are shown for a single-thread CPU job and GPU as a function of image pixel number. We present results for two different CPU+GPU sets.}
\label{tab:cpu_gpu_pol}
\normalsize
\end{table*}

\begin{figure}
\centering
\includegraphics[width=0.8\linewidth]{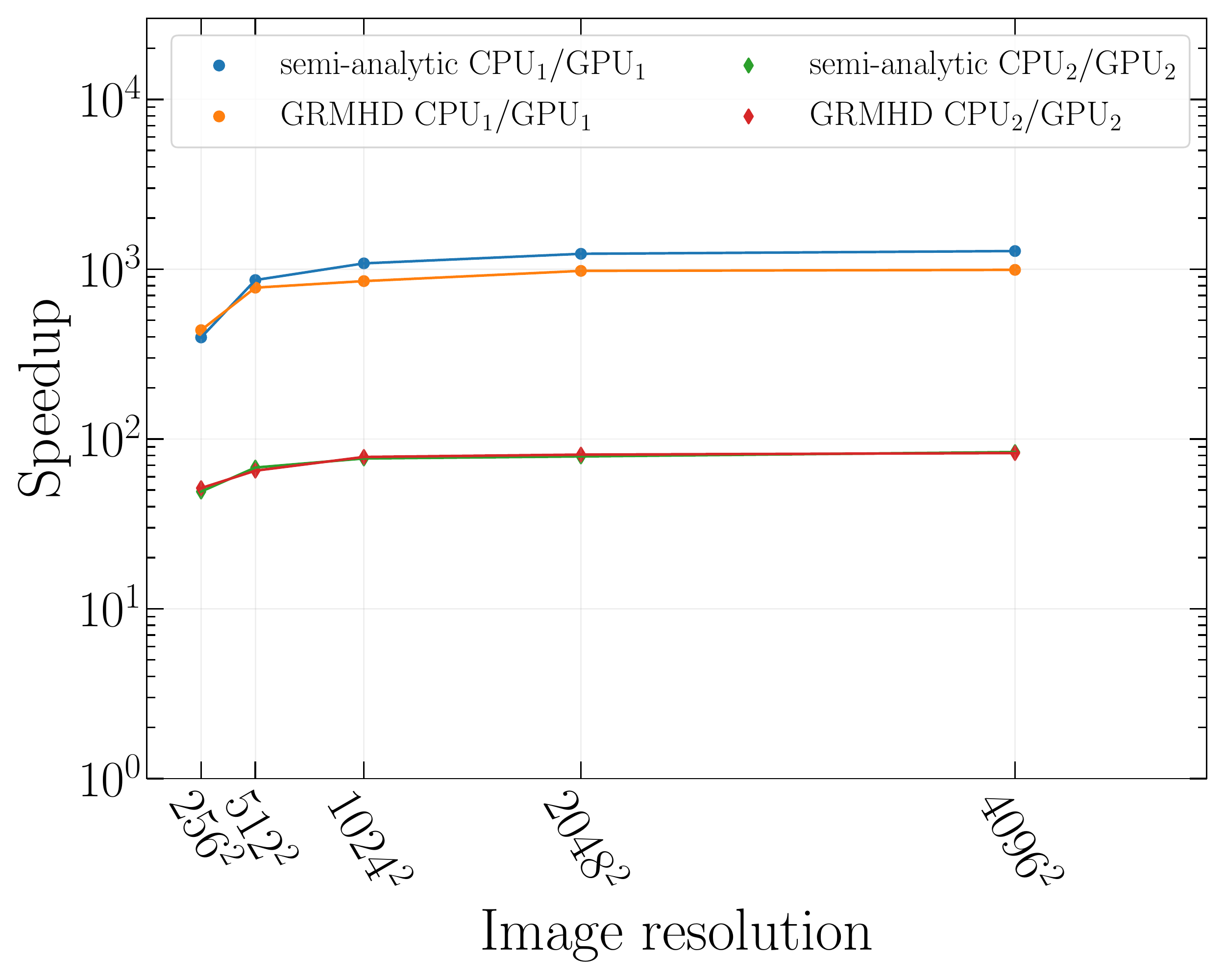}
\caption{GPU acceleration speedup factor as a function of image resolution. Results are shown for both types of models and both CPU+GPU sets. The CPU+GPU set No 2, compared to set No 1, has by design less double precision computing capability.}\label{fig:speedup}
\end{figure}

In this section we benchmark the GPU code against the single-thread CPU code. 
We carry out a speed test for both types of plasma models. All the tests are carried out for images including polarization for typical parameters (e.g., step size on geodesics) that are used when generating, e.g., the EHT image library.  
All speed tests assume fixed CUDA grid-block sizes with $16\times16$ threads. We carry out speedup tests using two sets of CPU+GPU. The first set is made of an AMD EPYC Rome (64c, 2.6GHz, 280 W) CPU and an Nvidia Ampere A100 GPU on installed on the European High Performance Computing (HPC) supercomputer. The second set is made of an Intel Core~i9(12th Gen Intel(R) Core(TM) i9-12900)  CPU and Nvidia GeForce RTX (3070 Lite Hash Rate)  GPU installed on an affordable desktop computer at home. 
Table~\ref{tab:cpu_gpu_pol} displays all run times of both codes in units of seconds for images with different resolutions. In Figure~\ref{fig:speedup} the numbers are compiled to show the speedup of the GPU code against the single-threaded CPU application. When run on HPC GPU, the code is up to 1279 and 980 times faster than the CPU code in the cases of semianalytic and GRMHD models, respectively. The speedup is also significant, of the order of ~100 for the second CPU+GPU set. The difference between the first and second GPU is double precision computing capability, which is lower by a factor of 30 for the second set\footnote{While the peak double precision performance for Nvidia A100 is 9.7 teraflops, it is only 318 gigaflops for the GeForce RTX.}. The best speedup results are obtained for images with high resolutions. Compared to the multithread CPU calculations (usually 16 or 24 threads) the speedup offered by the GPU is better, especially when using GPU with high double precision computing capability (our CPU code scales very well on a shared-memory node, hence the speed up compared to the multi-thread CPU run is $\sim$ speedup/N, where N is the number of memory-sharing threads). The behavior of the GPU code seems independent of the model type meaning that we are not strongly penalized for very frequent access of shared memory when imaging GRMHD simulations. Notice that semianalytic models do not use the managed memory at all as their structure is determined during radiative-transfer calculations.



\section{Discussion}\label{sec:discussion}

It is difficult to compare the performance of \texttt{cuda--ipole} to the GPU code GRay (developed by \citealt{chan:2013}); we solve fully polarized radiative transfer where we follow all four Stokes parameters while GRay solves the radiative-transfer equation for total intensity only (GPU versions of Odyssey and RAPTOR ray-tracing codes are also unpolarized, \citealt{pu:2016}, \citealt{davelaar:2018}). Also GRay is a single precision GPU code (private communications with the author). Additionally, the performance of the GPUs has improved since 2013. 

Our GPU code could be further sped up by another factor of 10 with adaptive ray tracing which reduces the number of light rays to follow \citep[following e.g.,][]{gelles:2021}. In principle a further code acceleration may be also obtained by switching to single precision calculations. However at some point the data coming in and out of the GPU takes as much time as the integration of radiative-transfer equations themselves. Hence there is a hardware limit for speedup. 

When using a GPU card with high double precision capability our code can be up to 50 times faster compared to a 24-thread CPU run. Although the power consumption of a GPU can be as high as 4 times that of the CPU alone, calculations using a GPU seem significantly greener ($\sim10$ times less energy consuming) even compared to a 24-thread CPU job. 

GPU-accelerated ray-tracing simulations of black hole images have several further advantages in context of EHT black hole image model fitting. For example, they enable dynamic model fitting using very complex models of accretion. For a single procedure of that sort we run the \ipole code up to 10,000,000 times. In the simple case of nonpolarized images of moderate (to low) resolution of $128\times128$, which takes roughly 10 s per image, this is equivalent to a total of 28,000 CPU thread hours computing time, or more than 10 days using 100 CPU cores (Yfantis et al. 2023, in preparation). This makes these runs computationally rather expensive. One has to be very careful and precise with the runs, making more exploratory works difficult. A speedup of the order of 1000 by using GPU computing power would decrease the computational time to several minutes, significantly enhancing the possibilities of Bayesian inference using GRMHD models. 

High-resolution images of accreting black holes, which can be now generated in reasonable time, enable more detailed studies of strong gravity effects that can be visible in both the total intensity and polarization of light of the photons or Einstein rings around the black hole shadow \citep[see, e.g.,][]{himwich:2020,alejandra:2021}. 

Finally, significantly accelerated ray-tracing radiative-transfer modeling may enable studies of emission from plasma with nonstandard electron distribution functions for which accurate prescriptions/approximations for synchrotron emissivities/rotativities do not exist and the integration of synchrotron emissivity over the electron energy distribution function requires numerical integration during ray tracing. This can be particularly useful when postprocessing general relativistic particle-in-cell simulations of particle acceleration around black holes \citep[see, e.g.,][]{ben:2021,ben:2022}.

\begin{acknowledgments}
We thank C.K. Chan for helpful comments and discussions. We also thank C.F. Gammie, H. Olivares, J. Vos and A. Jimenez-Rosales for discussions. We acknowledge support from Dutch Research Council (NWO), grant No. OCENW.KLEIN.113. This publication is also part of the project the Dutch Black Hole Consortium (with project number NWA 1292.19.202) of the research program the National Science Agenda which is financed by the Dutch Research Council (NWO). Part of the calculations have been carried out at Vega HPC at The Institute of Information Science in Maribor, Slovenia.
\end{acknowledgments}

%

\vspace{5mm}


\software{\ipole, python}







\end{document}